\documentclass[sigconf]{acmart}
\AtBeginDocument{%
  \providecommand\BibTeX{{%
    \normalfont B\kern-0.5em{\scshape i\kern-0.25em b}\kern-0.8em\TeX}}}

\copyrightyear{2024}
\acmYear{2024}
\setcopyright{rightsretained}
\acmConference[CHI EA '24]{Extended Abstracts of the CHI Conference on Human Factors in Computing Systems}{May 11--16, 2024}{Honolulu, HI, USA}
\acmBooktitle{Extended Abstracts of the CHI Conference on Human Factors in Computing Systems (CHI EA '24), May 11--16, 2024, Honolulu, HI, USA}
\acmDOI{10.1145/3613905.3650892}
\acmISBN{979-8-4007-0331-7/24/05}

\begin{document}

\title[Exploring the Impact of AI Value Alignment in Collaborative Ideation]{Exploring the Impact of AI Value Alignment in Collaborative Ideation: Effects on Perception, Ownership, and Output}

\author{Alicia Guo}
\email{axguo@mit.edu}
\affiliation{%
  \institution{Massachusetts Institute of Technology}
  \city{Cambridge}
  \country{USA}
}

\author{Pat Pataranutaporn}
\email{patpat@media.mit.edu}
\affiliation{%
  \institution{MIT Media Lab, Massachusetts Institute of Technology}
  \city{Cambridge}
  \country{USA}
}

\author{Pattie Maes}
\email{pattie@media.mit.edu}
\affiliation{%
  \institution{MIT Media Lab, Massachusetts Institute of Technology}
  \city{Cambridge}
  \country{USA}
}

\renewcommand{\shortauthors}{Guo, et al.}

\begin{abstract}
AI-based virtual assistants are increasingly used to support daily ideation tasks. The values or bias present in these assistants can influence output in hidden ways and affect how people perceive the ideas produced with these AI assistants, leading to implications for the design of AI-based tools. We explored the effects of AI assistants with different values on the ideation process and user perception of idea quality, ownership, assistant competence, and values present in the output. Our study tasked 180 participants with brainstorming practical solutions to a set of problems with AI assistants of different values. Results show no significant difference in self-evaluation of idea quality and perception of the assistant based on value alignment; however, ideas generated reflected the AI’s values and feeling of ownership is affected. This highlights an intricate interplay between AI values and human ideation, suggesting careful design considerations for future AI-supported brainstorming tools.
\end{abstract}

\ccsdesc[500]{Human-centered computing~Interaction design theory, concepts and paradigms}
\ccsdesc[500]{Human-centered computing~Empirical studies in interaction design}
\ccsdesc[500]{Human-centered computing~Empirical studies in HCI}
\ccsdesc[500]{Human-centered computing~HCI theory, concepts and models}
\keywords{Human-AI Interaction, Brainstorming, AI Assistants}


\begin{teaserfigure}
  \includegraphics[width=\textwidth]{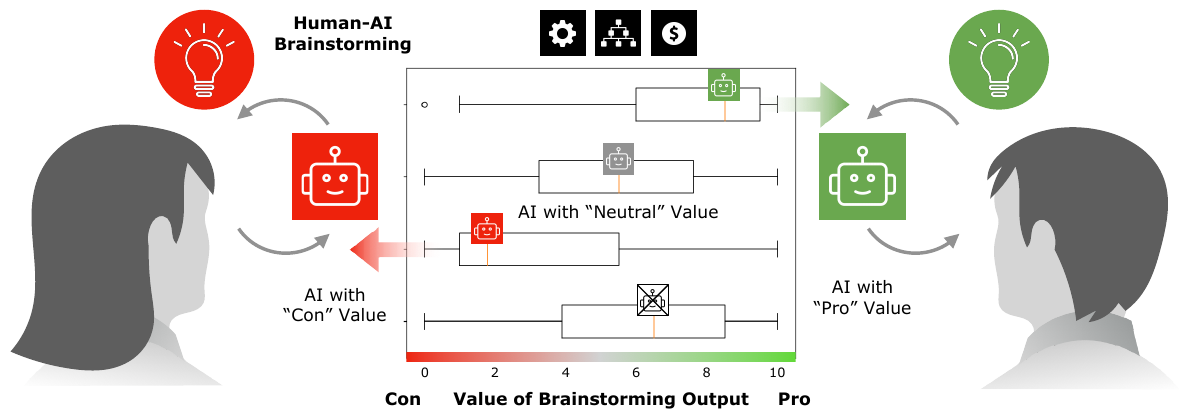}
  \caption{Users brainstorm with AI assistants exhibiting values either aligned with or contrary to their own.}
  \label{fig:cover}
  \Description{Two human silhouettes are shown brainstorming with three AIs exhibiting 'Con' (red), 'Neutral' (gray), and 'Pro' (green) values, represented by robot faces. A central horizontal bar graph shows the AI-influenced brainstorming output ranging from 'Con' to 'Pro' values. Gear, currency, and hierarchy icons indicate the brainstorming domains.}
\end{teaserfigure}


\maketitle

\section{Introduction}

The integration of large language models (LLMs) into everyday tasks, such as writing, programming, and customer service, is becoming increasingly common \cite{Ippolito2022CreativeWW, code, Li2022CompetitionlevelCG}. With the release of GPT-4, AI assistants are being further embedded into the workflows of common tasks. Though this technology can increase productivity and convenience,the long-term effects of these new work patterns on individuals' output and their sense of task ownership and agency remain unknown.

Specifically, the increasing prevalence of AI systems as creative partners for tasks like brainstorming and inspiration \cite{Kim2023TheEO, lin2020your, mayAI} raises important questions about its impact on individuals' sense of ownership and agency during the ideation process \cite{koch_agency_2021, buschek_nine_2021}. Collaborative creation between humans and AI assistants represents a complex interplay of factors rooted in autonomy, transparency, and alignment of values. The impact of alignment or misalignment of values (the core principles or beliefs that guide behavior) between humans and AI systems on human-AI co-creation is still an under explored area. As AI capabilities continue to advance, it is critical to identify connections between the process of ideation and creation, especially with respect to ownership, existing applications of AI for ideation, and considerations of bias in algorithmic systems. 

Through an assisted brainstorming experiment, we explore how individuals' interaction with AI assistants that are customized to hold different views (the same as the individual, opposite to theirs, or neutral) affects creative ideation. We create brainstorming prompts to elicit participant views in three areas (economics, automation, organizational hierarchies). Participants answered two prompts, one where they worked alone and the other where they were paired with an AI assistant exhibiting a value, or bias that may or may not differ from the participants' own, as assessed in a pre-experiment participant survey. 

Our analysis reveals that the participants' final ideas integrated and reflected the values displayed by the AI assistants they conversed with during ideation. However, when the AI assistant's values were misaligned with the participant's own views, users reported increased difficulty and frustration. Importantly, across all conditions, \textit{working with the AI assistant diminished the participants' reported sense of ownership over the ideas produced compared to brainstorming alone without AI assistance.} These findings point to subtle influences whereby AI systems shape ideation outputs and reduce feelings of creative agency. Though this study has limitations in it's scope and constraints, this suggests implications for the design of human-AI collaborations, highlighting the need to balance autonomy and alignment in co-creation and the need for careful development of the AI assistants used in these tools.

\section{Related Works}

An important step in creative problem solving is the ideation process, where ideas are produced prior to the process of selecting ideas \cite{joyner}. A distinction can be made between divergent processes, where one generates multiple ideas and convergent processes, focusing on identifying a single, optimal solution. One model for the creative problem solving process consists of the three stages of problem finding, problem solving, and solution implementation, where each step involves independent "ideation-evaluation" substeps \cite{Basadur1982}. Ideation here refers to the divergent process of generating ideas and reserving judgement for later in the evaluation process where ideas are narrowed down to the best ones \cite{Getzels1975ProblemFindingAT}. 

\textit{AI tools for creation and assistance.} Large language models (LLMs) are a class of machine learning models trained on text data. Recent advances in natural language processing have led to the development of larger and more powerful LLMs \cite{brown2020language, NIPS2017_3f5ee243}, such as Claude and GPT-4 that are pre-trained and able to perform well without much downstream fine-tuning or training, instead relying on prompts or chatting in natural language interactions. They have increasingly become a part of our lives embedded in tools for various tasks such as creative writing assistants \cite{Ippolito2022CreativeWW}, inspiration for scientific writing \cite{spark}, tools for code generation \cite{code, Li2022CompetitionlevelCG}, in home assistants that are voice-based such as Alexa, chatbots for customer service \cite{Maedche2019AIBasedDA}, and much more.

Several works study AI-based digital assistants' effects in ideation processes \cite{Kim2023TheEO}, including in dance improvisation \cite{viewpoints}, design ideation \cite{lin2020your}, and drawing sketches \cite{oh2018lead}, and creating mood boards \cite{mayAI}. LLMs were found to be especially useful for generating ideas during the ideation stage, even more so when users had trouble coming up with ideas on their own. Novel ideas can be suggested even if low quality, and require iterative prompting to create better outputs \cite{wan2023it}. Intelligent agent systems trained for the brainstorming process are found to be effective in introducing new topics with users perceiving them to be comparable or even more capable than humans \cite{Wang2018DesignOA}. AI can improve idea generation, however the effectiveness can also depend on user's attitudes towards AI \cite{Chiu2023InvestigatingTI}.

\textit{AI authorship}. Work has been done examining language models' capabilities in creative and argumentative writing show GPT-3's abilities and consider metrics for good collaboration including writers' feeling of ownership \cite{coauthor}. Other investigations into writers' feelings of authorship could be effected by the interaction (suggestions vs. direct generation) \cite{lehmann}. 

\textit{Values in AI}. LLMs have been shown to carry the biases of their training data, with consequences in exacerbating societal biases in areas like hiring, lending, content moderation, and healthcare \cite{Ferrara2023ShouldCB, Maedche2019AIBasedDA}. A previous study investigates whether using a biased language model-powered writing assistant affects what users write and changes their opinions \cite{Jakesch_2023}. The study asked participants to respond to an argument about social media ("Is Social Media Good for Society?”) with an AI-powered text completion interface. Some users got suggestions from a model biased to argue social media is good, others from a model arguing it is bad. Participant responses were more likely to contain the opinion supported by the biased model they interacted with. The biased models also shifted participants' attitudes in a later survey about social media, suggesting an actual opinion change beyond just conveniently accepting suggestions. The authors argue this demonstrates a new paradigm called "latent persuasion" where language models shift users' views and writing by making some opinions easier to express than others \cite{Jakesch_2023}. Our work investigates participants' evaluations of idea quality, ownership, and ideation process in addition to the output to examine the effects of value alignment between human and AI in these situations on human perception of the ideation process.

\section{Methodology}

\subsection{Study Overview} 

We examine whether conversing with AI assistants during a brainstorming task impacts people's perceptions of the ideation process itself, their sense of ownership over the generated ideas, the presence of the AI input in the final idea, as well as the effects of personal value alignment with the AI. We conducted an online experiment asking participants (N=180) to brainstorm practical solutions to a set of problems with the AI assistants exhibiting different values along three domains: Economics, Automation, and Organizational Hierarchy. We use \textit{values} to refer to the explicit principles that guide behavior and \textit{views} to refer to the stances formed from these values. \textit{Bias} refers to an inclination for or against these values or views.

The research questions and methodology has been pre-registered as "Brainstorming Study - AI Values" with protocol number \#141856 via https://aspredicted.org/. The study was conducted with a total of 180 participants (60 per domain) from Prolific, with the experiment survey results collected through Qualtrics. The sample was made up of USA-based participants ages 18-60 with 24.5\% aged 18-24, 31\% aged 25-34, 21.7\% aged 35-44, 14.7\% aged 45-54, . 46.7\% self-identified as female, 47.2\% identified as male, 1.6\% identified as non-binary, and 2.2\% preferred not to say or to self-describe.

\begin{figure*}[!h]
\centering
\includegraphics[width=1\textwidth]{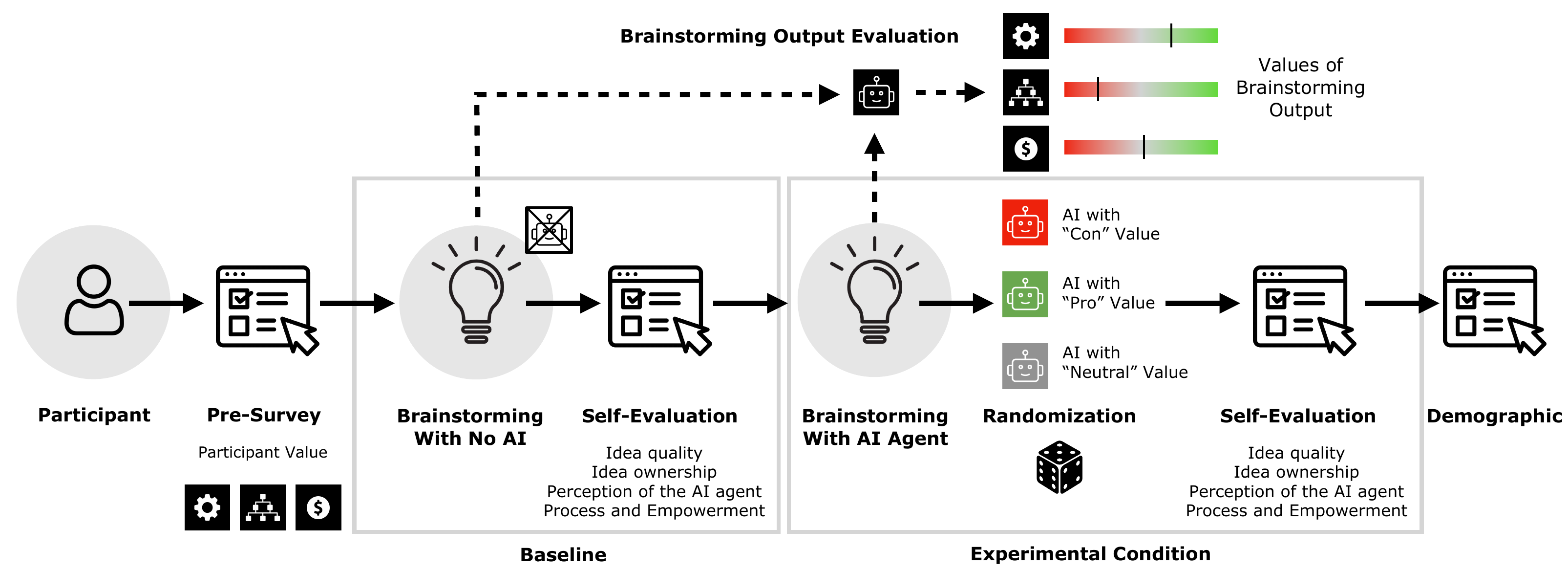}
\caption{Overview of the study steps each participant experiences. Participant: (1) indicates personal values in the pre-survey, (2a) performs a brainstorming task individually and evaluates the experience, (2b) performs a brainstorming task with an AI assistant (randomly chosen from \textit{Pro}, \textit{Neutral}, or \textit{Con} AI conditions) and evaluates the experience, and (3) answers a post-survey on study experiences and demographics.}
\Description{A flowchart showing a study process which begins with a participant taking a pre-survey, followed by an individual brainstorming task, then a self-evaluation (random order). After being randomized to brainstorm with an AI with 'Con,' 'Pro,' or 'Neutral' values, participants evaluate again and provide demographic information. The top right shows the final output values ranging from 'Con' to 'Pro'.}
\label{fig:study}
\end{figure*}
\subsection{Value Domains} \label{domains}

We conducted this study with three distinct value domains: Economics (economic ideology), Automation (perspective on automation), and Hierarchy (perspective on organizational structure). Each domain represents a topic for which we can define opposing values, with a \textit{Pro} side on one end and a \textit{Con} side on the other. We keep a clear dichotomy for each domain which does not take into account all of the nuances of the values, but are distinct enough to reveal the values of the users. These domains were chosen from a pre-experiment survey on multiple subjects that could have elicited a wide variety of views - we chose the three domains that had sufficient views on both ends of the domain (Table~\ref{tab:domains}). For each of the value domains, two brainstorming prompts were created and randomly assigned to the brainstorming sessions with and without AI assistants. These were designed to allow users to include elements of the value of the domain being studied in the implementation of the solution without having it be the sole focus of the session (full prompts in Appendix \ref{appendixA_questions}). 

\noindent \textbf{Automation}: Measures attitudes on automation and technological disruption of work, simplified into two perspectives; one view focuses on pros of automation, e.g., efficiency and progress, and the other focuses on its cons, e.g., labor replacement. 

\noindent \textbf{Economics}: Measures the economic leanings (from laissez faire capitalism to socially progressive) of a person using descriptions from the Economic axis of the 8 Values quiz \cite{8values} to produce market (\textit{Con}) and equality (\textit{Pro}) views.

\noindent \textbf{Hierarchy}: Measures preferences for organizational structure, simplified into a domain from hierarchical (pro) to flat structures (con).

\begin{table}[!h]
\begin{center}
\begin{tabular}{| p{0.17\linewidth} |p{0.35\linewidth} | p{0.35\linewidth} |}\hline
\textbf{Domain} & \textbf{Con (0)} & \textbf{Pro (10)}\\\hline
Economics & I support rapid growth, laissez-faire capitalism, lower taxes, deregulation, and privatization. & I support even value distribution, equality via progressive tax, and social programs.\\\hline
Automation & I value job retention and meaningful work through human labor. & I support automation for cost-effectiveness and efficiency.\\\hline
Hierarchy & I support flat organizations for promoting equality and fostering collaboration. & I support hierarchical structures for clear roles and streamlined decision-making.\\\hline
\end{tabular}
\end{center}
\caption{\textit{Pro} and \textit{Con} value statements for each domain.}
\label{tab:domains}
\end{table}

\subsection{Study Procedures}

To examine the effects of personal vs AI values (whether the AI’s values align with or oppose the user’s), participants were assigned to a random value domain and first indicated their own values on the given value domain using a scale of 0-10, where 0 is \textit{Con} and 10 is \textit{Pro}, in a single question with descriptions from Table~\ref{tab:domains} (distribution shown in top row of Figure~\ref{fig:gpt-split}). Each participant then completed two timed brainstorming sessions: one independently and one paired with an AI assistant, as shown in Figure~\ref{fig:study}. These were completed in a random order. For each brainstorming session, participants were given a real-world situation and asked to suggest a solution either by themselves (\textit{No AI}) or in collaboration with the AI assistant after chatting. Participants were randomly assigned to one of three AI bias conditions for their brainstorming session: \textit{Pro}, \textit{Con}, or \textit{Neutral.} Participants in the \textit{Pro} group brainstormed with an AI assistant displaying values on the \textit{Pro} end of the value domain. Participants in the \textit{Con} group brainstormed with an AI displaying values on the opposite \textit{Con} end of the value domain. Participants in the \textit{Neutral} group brainstormed with an AI displaying balanced values in the middle of the two ends of the value domain. The details of the values are expanded upon in section \ref{domains}. Participants were first given time to either chat with the AI or think by themselves before the text box for the final idea response appeared in order to ensure they engaged with the AI assistant in the process, then asked to write at least four sentences in their final response.  After each session, participants completed survey measures assessing perception of idea quality, ownership over the idea, perception of the AI assistant (when applicable), and the process of the ideation experience. 

We aim to answer the following research questions that informed the design of the experiment:

\begin{itemize}
    \item How will the values of the AI that the user brainstorms with (Pro, Neutral, Con) affect \textit{the values present in the final idea}?
    \item How will the value alignment of the AI with the user (same as user, opposite of user, or neutral) affect \textit{self-evaluation and external evaluation of idea quality}, \textit{self-evaluation of idea ownership}, \textit{user perception of the AI assistant}, and the \textit{user’s perception of the brainstorming experience}?
\end{itemize}



\subsection{AI assistant Interaction}

For the study, we developed AI assistants exhibiting perspectives in the value domains using a prompting approach with the GPT-4 language model. A custom interface was created for users to interact with this assistant in the style of common chat interfaces. For \textit{Pro} value assistants and \textit{Con} value assistants the prompt followed the template "You are a brainstorming partner. You support [\textit{value description}]. Reply in 30 words or less." For \textit{neutral} value assistants the prompt followed the template "You are a brainstorming partner. You are a neutral party between the sides of [\textit{Pro} value description] and [\textit{Con} value description] and see the benefits and consequences of both sides." We found this to be effective in responding to questions with suggestions that aligned with the prompted value. Separate models were constructed for each of the three domains and AI conditions of \textit{Pro}, \textit{Neutral}, and \textit{Con}. The sampling temperature was set to 0.7 to create more variation in responses. An example interaction is shown in Appendix \ref{appendixE_example}.

\subsection{Measurements}

Initial values of the user were measured on a self-scored scale of 0-10 as detailed in the values Table ~\ref{tab:domains}. After each brainstorming session, users answered a set of questions providing their self-evaluation of idea quality, self-evaluation of idea ownership, perception of the AI assistant (if applicable), and self-evaluation of brainstorming process and empowerment on a 5-point Likert scale. Survey questions are detailed in Appendix \ref{appendixB_evaluation}.

\subsection{Analysis}

We analyzed outcome variables using descriptive statistics and ANOVA models comparing the experimental conditions and control condition. For each measure, we analyzed the individual and composite scores associated with each scale. We sampled down the responses for the No AI condition to 20 to match the size of the Pro, Neutral, and Con conditions. 

Statistical tests were used independently for each separate Likert question. We first assessed if the normality assumption was met for each distribution using the Shapiro-Wilk test. If the normality assumption was not met, we performed a Kruskal-Wallis test followed by a post-hoc Dunn test using the Bonferroni error correction. If the normality assumption was met, we then conducted a homogeneity test using a Levene test to assess whether the samples were from populations with equal variances. If the samples were not homogeneous, we ran a Welch analysis of variance (ANOVA) and a Tukey post-hoc test. If the samples were homogeneous, we ran a basic ANOVA test. 

Additionally, the final ideas were evaluated by a prompt crafted to allow the GPT-4 natural language model to analyze the semantic content of participants' ideas. We generated scores reflecting the presence of relevant value domain in users' brainstormed ideas. While not at the level of human rater sophistication, this computational analysis complemented the subjective self-report measures by providing an objectively derived evaluation of how values were reflected in the ideation content and has been shown to be a good general model for sentiment analysis \cite{amin2023affective}. Each evaluation was done with the prompt "Sentences with a high [domain] score support [\textit{Pro} value description]. Sentences with a low [domain] score support [\textit{Con} value description]. Based on a given paragraph, output a number from 0 to 10 on how present the [domain] value is. 0 for [\textit{Con} value], 10 for [\textit{Pro} value description]." Two responses without a score were not included because the participants did not complete their response. We recruited 3 human evaluators to validate these scores. They rated a sample of half of the Economic value responses and we found the correlation coefficient between human and GPT-4 scores to be r = 0.8375.

\section{Results}

\subsection{Values in Brainstorming Output}

\begin{figure*}[!h]
\centering
\includegraphics[width=1\textwidth]{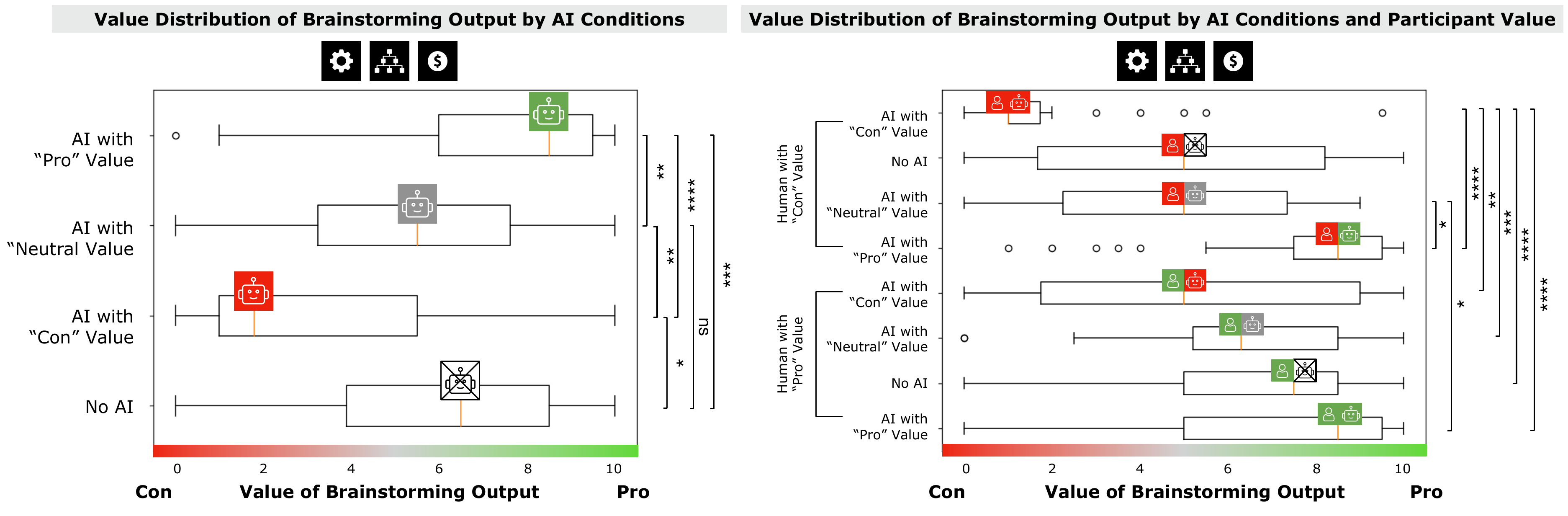}
\caption{Output values split by AI value and participant values}
\small{\textbf{(Left)} GPT-4 evaluations of participant responses, split by AI condition: "Pro" (green robot), "Neutral" (gray robot), "Con" (red robot), or no AI (robot with a cross). Responses aligned with the respective AI's values. \textbf{(Right)} GPT-4 evaluations split by both human and AI values. Top: "Con" human values (red person), bottom: "Pro" human values (green person), each split by AI condition. Graphs show min, Q1, median, Q3, and max values per condition.}

\Description{Two box plots showing evaluations of values in responses averaged across all domains. The x-axis scale shows 0-10 (from Con to Pro) for the Value of Brainstorming Output. The left box plot titled "Value Distribution of Brainstorming Output by AI Conditions" has boxes for AI with "Pro" Value, AI with "Neutral" Value, AI with "Con" value with decreasing means and similar ranges. The No AI box has a mean between "Neutral" AI and "Pro" AI. For the right box plot titled "Value Distribution of Brainstorming Output by AI Conditions and Participant Value", each box is further split by the human value self-ranking ("Pro" or "Con") for 8 boxes of each human value and AI condition pair. }
\label{fig:gpt-overall}
\end{figure*}

We first look at the values present in the brainstorming responses for all three domains. We found a significant difference in how the three AI conditions \textit{Pro}, \textit{Neutral}, and \textit{Con} affected the values present in the final brainstormed idea compared to each other and the \textit{No AI} condition. A one-way ANOVA found a significant difference between conditions (F(3, 219)=14.03, p=2.2E-8).

Post-hoc Tukey tests reveal responses in the \textit{Pro AI} condition rated significantly higher on the value dimension (M=7.49, SD=2.76) than the \textit{Neutral AI} condition (M=5.51, SD=2.97; p=.005),  the \textit{Con AI} condition (M=3.65, SD=3.45; p < .001), and the \textit{No AI} condition (M=5.91, SD=3.16; p=.039). The \textit{Con AI} condition also had significantly lower ratings than the \textit{Neutral AI} condition (p=.01) and the \textit{No AI} condition (p=.001). However, no statistically significant differences were present between the \textit{Neutral AI} and \textit{No AI} control conditions (p=.907), suggesting that the presence of an AI assistant in the brainstorming process did not impact the value evaluation. The trends show that people’s responses tended to align with the values of the AI assistant they brainstormed with (Figure~\ref{fig:gpt-overall}).

When we further split these groups by the personal values of the participant, the data violated normality assumptions according to a Shapiro-Wilk's test. Thus, Kruskal-Wallis tests were conducted, using post hoc Dunn's with Bonferroni corrections to determine if there were significant differences between groups. We observe a trend that when participants were brainstorming with an AI assistant of similar values, this further amplified the value present in the response by shifting the mean and medians of the value. On the other hand, we observe the trend that brainstorming with an AI assistant of differing values pulls the value present in the response closer towards neutral.

\begin{figure*}[!h]
\centering
\includegraphics[width=1\textwidth]{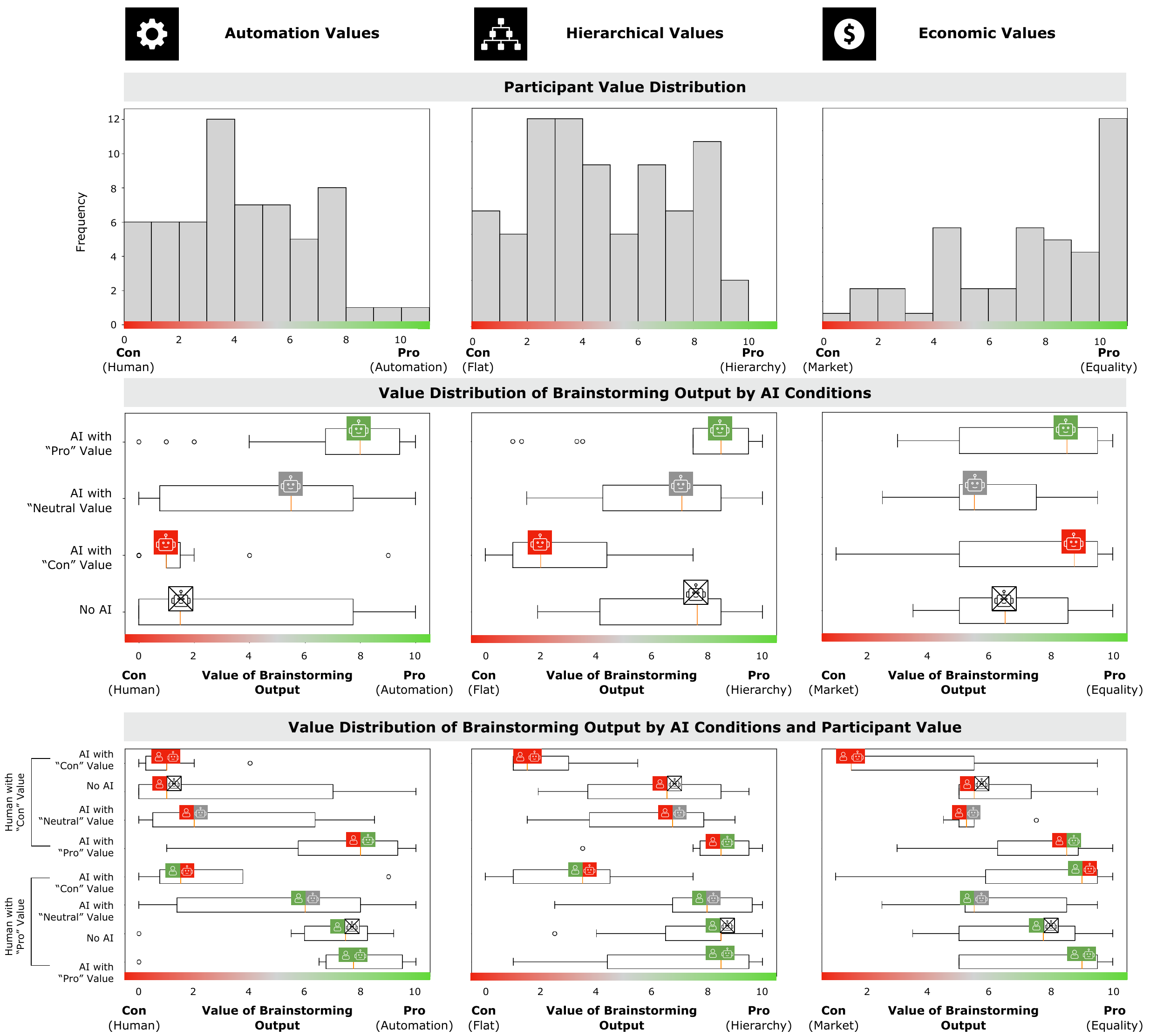}

\caption{Output values further split by domain}

\small{\textbf{(First row)} Participant value distributions for Economic, Hierarchy, and Automation domains from the pre-survey. \textbf{(Second row)} Response value scores by AI condition for each domain: "Pro" (green robot), "Neutral" (gray robot), "Con" (red robot), no AI (robot with cross). 
 \textbf{(Third row)} Response value scores by human and AI values. GPT-4 evaluations split by human values: "Con" (red person, top), "Pro" (green person, bottom), and AI condition shown by "Pro" (green robot), "Neutral" (gray robot), "Con" (red robot), no AI (robot with cross).}

\Description{There are 9 graphs arranged in a 3x3 grid. The left column has Automation domain graphs, the middle column is Hierarchical, and the right column is Economic. The first row shows bar graphs of participant self-rankings for the respective value on the "Con" to "Pro" spectrum. The second row shows "Value Distribution of Brainstorming Output by AI Conditions" box plots of the value evaluations for each domain value split by AI conditions "Pro", "Neutral", "Con", and no AI. The third row shows "Value Distribution of Brainstorming Output by AI Conditions and Participant Value" of responses from the second row, split further by human value "Pro" or "Con" and the 4 AI conditions for a total of 8 boxes per graph. The boxes show a trend where the "Con" human value with the "Con" AI had the most "Con" values present and the "Pro" human value with the "Pro" AI had the most "Pro" values present.} 
\label{fig:gpt-split}
\end{figure*}

Taking a deeper look into the distributions when separated by value domain (Figure~\ref{fig:gpt-split}), we see the same trend for the Automation and Hierarchy domains. The initial distribution of participant values for the Economic domain is skewed toward the \textit{Pro} side, which could explain the lack of difference in the response values between the \textit{Pro AI} and \textit{Con AI} conditions. When we look at the Economic response values further split by all human alignment and AI alignment values, we see an interesting phenomenon where the median and mean scores of the participant with \textit{Pro} values brainstorming with the \textit{Con AI} assistant skew towards \textit{Pro}. This could be explained by the initial skew of the participant values, where many ranked themselves strongly on the \textit{Pro} side.

\subsection{Self-evaluation questions}

For the self-evaluation of idea quality, self-evaluation of idea ownership, and self-evaluation of brainstorming process and empowerment question groups, we averaged the question responses together to form overall \textit{idea quality}, \textit{idea ownership}, and \textit{process} scores. 

\begin{figure*}[!h]
\centering
\includegraphics[width=1\textwidth]{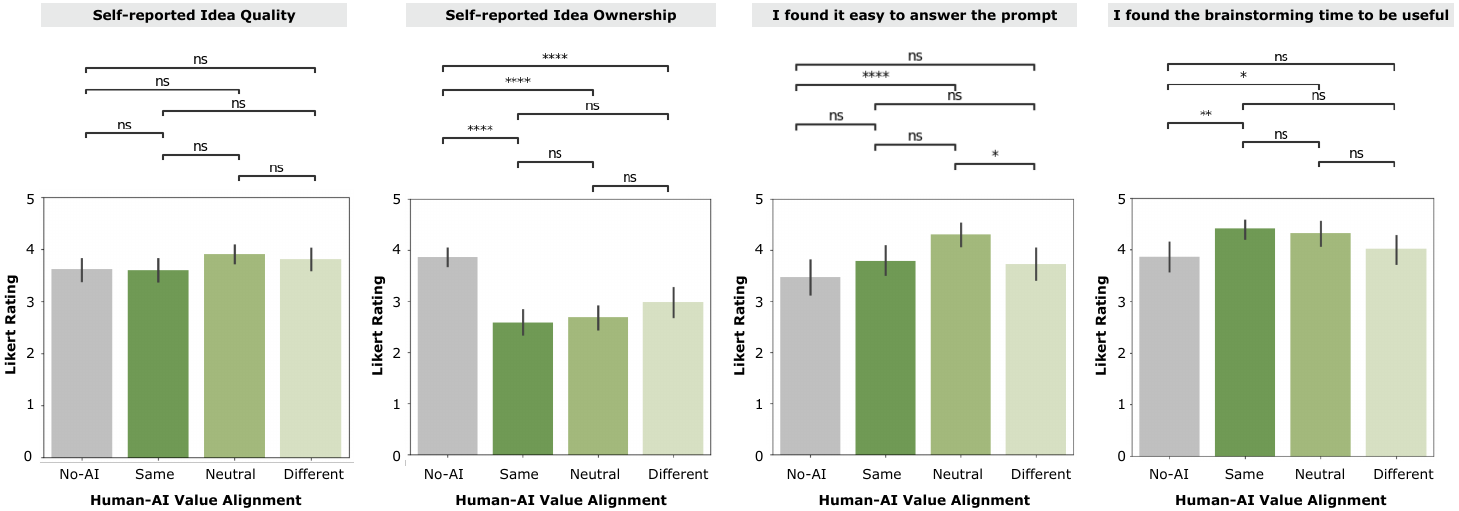}
\caption{Significant post-survey questions across human-AI conditions }
\small {1) averaged idea quality score 2) averaged ownership score 3) "I found it easy to answer the prompt" from the process questions 4) "I found the brainstorming time to be useful" from the process questions. }
\label{fig:averages}
\Description{The graphs depict participant answers for the following: Average idea quality score, Average ownership score, Responses to the process question: "I found it easy to answer the prompt", Responses to the process question: "I found the brainstorming time to be useful"}
\end{figure*}

\noindent \textit{Human-AI value alignment and self-evaluation of idea quality} \newline
A one-way ANOVA found no significant difference (F(3, 240)=2.16, p=0.094) between human-AI value alignment conditions (No AI (M=3.62, SD=0.84), Same (M=3.6, SD=0.92), Neutral (M=3.91, SD=0.7), Different (M=3.81, SD=0.79)) and self-reported idea quality questions (first graph of Figure~\ref{fig:averages}).

We found statistically significant differences in human-AI value alignment (No AI, Same, Neutral, Different) and averaged self-reported idea ownership.
A one-way ANOVA found a significant difference between conditions (F(3, 60)=22.53, p=7.3E-13). Post-hoc Tukey tests reveal the presence of AI significantly reduced the idea ownership score compared to the No AI condition (M=3.87, SD=0.69); Same condition (M=2.59, SD=0.99; p < .0001), Different condition (M=2.98, SD=1.10; p < .0001), Neutral condition (M=2.68, SD=0.99; p < .0001). We found no evidence that the different AI values had an effect on self-evaluation of idea ownership, which aligns with the lack of statistically significant difference to the ratings on "The answer was influenced by the assistant’s responses."
Looking more closely at the individual questions, "The answer was fully my idea" and "When I think about it, I see a part of myself in the answer" yielded statistically significant differences between conditions in a similar pattern of participants feeling more ownership of their idea in the No AI condition compared to the AI conditions. 

\textit{Human-AI value alignment and process empowerment} \newline
We found statistically significant differences in human-AI value alignment (No AI, Same, Neutral, Different) for the "I found it easy to answer the prompt" and "I found the brainstorming time to be useful" questions.
Participants found it easier to answer the brainstorming prompt with the Neutral AI assistant compared to having an AI assistant with opposing values or brainstorming without an assistant. A one-way ANOVA showed a significant difference between conditions (F(3, 60)=5.62, p=0.001). Post-hoc Tukey tests reveal the Neutral condition (M=4.31, SD=0.85) rated significantly higher scores than the Different condition (M=3.73, SD=1.17; p=0.037) and the No AI condition (M=3.47, SD=1.31; p < 0.0001). A one-way ANOVA showed an overall difference between conditions (F(3, 60)=4.37, p=0.005) with post-hoc Tukey tests revealing that participants found brainstorming time in the Neutral AI (M=4.33, SD=0.89; p=.043) and Same AI (M=4.41, SD=0.75; p=0.010) conditions significantly more useful than not having an AI assistant (M=3.87, SD=1.13). 

\textit{Human-AI value alignment and the perception of the AI assistant} \newline
A one-way ANOVA found no significant difference (F(3, 60)=2.16, p=0.094) between human-AI value alignment conditions (No AI (M=3.62, SD=0.84), Same (M=3.6, SD=0.92), Neutral (M=3.91, SD=0.7), Different (M=3.81, SD=0.79)) and self-reported perception of AI assistant questions.

\subsection{Moderator Analysis}
For the total 180 responses, we ran a linear regression to explore how moderators could effect the outcome variables, with figures and full details in Appendix \ref{appendixC_analysis}. Linear regression analysis found a significant positive correlation between ownership and process scores (r=0.318, p < 0.001) with a slope coefficient of 0.243 (SE=0.054) suggesting a positive relationship between how much ownership the users felt to how positively they viewed the brainstorming process. Linear regression analysis also found a significant positive correlation between idea quality scores and process scores (r=0.289, p < 0.001) with a slope coefficient of 0.104 (SE=0.026) suggesting a positive relationship between how positively users viewed the quality of their ideas to how positively they viewed the brainstorming process.

\subsection{Brainstorming feedback}
Participants' brainstorming sessions averaged 10.3 messages per session (including the AI assistant) with a standard deviation of 3.1 messages. We then asked participants to tell us about their experiences in the brainstorming sessions. 

Most of the feedback mentioned the AI being helpful to the process, with a few who did not feel familiar with the brainstorming prompt relying more on the AI. Participants thought that the "\textit{AI came up with good ideas quickly}," "\textit{was very helpful}," "\textit{made the brainstorming more efficient}" and was "\textit{easy to talk to and bounced well of my messages and ideas.}" Some participants wrote that they "\textit{prefer[red] to talk with the AI [assistant], it made it easier to brainstorm.}" Within this group, people used the AI assistant differently; some used the assistant to expand on their ideas, with another writing that "\textit{the AI agent definitely helped me "kickstart" my ideas. While it didn't write anything for me, it did help me focus on a specific topic for the prompt given.}" 

Amongst the users that did not find the assistant to be helpful, some participants felt that the AI's responses were not knowledgeable enough or specific enough to be helpful, that "\textit{the AI's ideas didn't really align much with my vision}," while others felt the responses to be too short or the AI to be too stuck in its suggestions. Depending on how people interacted with the assistant, they could end up in a situation where "\textit{parrotted back my own idea without helping me to develop it very much}" or where "\textit{the AI agent had pretty generic responses, so that wasn't much help.}" while it was mentioned that participants took many iterations to get to where they wanted to be on the idea.

Familiarity seems to play a role in how much users took the AI's suggestions. A few mentioned that they deferred to the assistant's ideas because they lacked familiarity with the topic with one participant with 19 years of experience in the field of the prompt expressing frustration that the "\textit{AI agent I didn't enjoy that much ... I do agree with the AI that replacing some repetitive tasks with AI would be great, but overall wasn't my solution.}" Another said "\textit{I didn't find the AI agent helpful other than it greatly annoyed me}" when it suggested ideas that the user didn't agree with.

Even for the participants who found the AI assistants helpful, a few found their sense of ownership diminished: \textit{"It was easier to think when the AI agent helped but I did not feel like it was totally my idea so I didn't feel as proud of the answer as I did on the first task."} and \textit{"this saved thinking time on my end, but the idea didn't feel original nor like I came up with it on my own."}

The range of experiences shows a highly varied vision of how users interacted with the AI and that what one person might find helpful, another might not. 

\section{Discussion}

\subsection{Implications}

Our study suggests many considerations for the design of human-AI interactions. We found that\textit{people interacting with AI assistants for problem solving can be subtly influenced by the AI's values.} These values can appear in the participants' brainstorming output without them feeling influenced by the interactions, either amplifying the users' own values or pulling them more towards the center. This shows the potential for LLMs to influence user outputs, suggesting careful design is required when incorporating AI assistants for ideation tasks. This could come in the form of more transparency in how models are trained or how outputs are selected, offering a wider range of outputs across multiple perspectives, or more policy towards education for the general user of AI applications on how to be conscious of their use of AI \cite{pataranutaporn2023influencing}.

Across all conditions, interacting with an AI assistant significantly reduces the users' sense of ownership over their ideas compared to ideating alone,\textit{ }illustrating \textit{a potential trend of AI diminishing perceptions of creative agency} that must be further explored. Perception of ownership and idea quality, but not how influenced the user felt, were associated with a better brainstorming process. Perceived influence and actual value change had no correlation, suggesting that \textit{people might not be able to accurately perceive how influenced they were in the human-AI interaction.} This could be affected by the brief period of brainstorming given to participants - possible design mitigations could include a focus on guiding questions as opposed to solutions or interfaces that allow for easy iterative integration of ideas from both human and AI. 

While the participants found the AI to be helpful during brainstorming, they perceived no difference in their evaluation of the quality of the idea. Users found it much easier to brainstorm in sessions with the Neutral AI assistant as opposed to one holding views differing from their own or the No assistant condition. They also found the brainstorming time with the AI assistant of similar views to themselves and the Neutral AI assistant to be more useful, suggesting that brainstorming with the Different AI assistant condition added friction. In a few cases when the AI assistant exhibited values opposing the user's own, participants reported increased difficulty and decreased usefulness during ideation in the feedback. This friction suggests that \textit{aligning AI assistant views with human creators may be critical for fluid co-creativity; on the other hand, designers need to avoid creating "echo chambers" that reinforce human biases if alignment is not carefully considered.} Examining the long-term effects of interacting with AI assistants on human decision-making and bias perception would provide deeper insights into the evolving relationship between humans and AI. The question it leads to is how to design AI assistants and systems that accurately portray or increase sense of ownership without becoming echo chambers.

An intriguing application of our findings is in the calibration of AI biases to counteract human biases in decision-making processes. Tuning the bias of AI assistants to counterbalance human bias could be an avenue toward more equitable and fair outcomes (e.g., loan approvals, looking at job applications) \cite{viegas2023system}. This approach would necessitate a careful study of the ethical implications and the mechanisms for implementing such balance in AI systems, ensuring that AI assistance supports humans without undermining their autonomy or perpetuating systemic biases.

\subsection{Limitations}

There are many factors that could affect an experiment like this. This experiment was conducted in a Qualtrics survey with crafted brainstorming prompts that users had no personal stake in. Additionally, users mentioned that they felt that five minutes (with additional grace time to finish responses) was too short (averaging 10.3 messages), and they may have chosen to follow the AI's suggestions due to that. Our study consisted of a single AI brainstorm session per participant, so we were not able to assess how each participant would have behaved and been affected in all of the AI conditions. We chose to use one self-scored question as it was the simplest after testing out multiple different measures, but real world values are much more nuanced and do not exist on a single dimension. The feedback from participants reveals insight into the motivations for their evaluations and a study with a supplementary interview could potentially offer explanations for the relationships between variables.

\subsection{Future Work}

This study points to several possible directions of continued investigation regarding bias and values in human-AI ideation processes. Future research could explore the nuanced dynamics of human-AI ideation in various contexts. For example, studies could investigate how different types of biases in AI assistants influence decision-making in critical domains like healthcare, legal judgment, or policy formulation. 

Addressing the limitations, a potential experiment to mitigate the effects of participants not having a personal stake in the brainstorming task would be to conduct the study in an untimed, real-world workshop with a prompt relevant to a specific community, with evaluations done by external raters. We would also like to explore longer brainstorming sessions to see whether user's values have changed over extended use of biased AI assistants. 

Other important directions would be to explore what happens when a user brainstorms with multiple assistants with opposing values during a session \cite{pataranutaporn2021machinoia}, what a more customized UI or assistant could do to improve the process, or when the AI system proactively asks questions \cite{danry2023don} and makes suggestions to guide the ideation process, and how user perception of ownership and agency affects the life of the brainstormed idea post-study. We hope that the findings would further define design guidelines for incorporating AI into ideation processes.

\section{Conclusion}
The increasing prevalence of AI systems as creative partners for collaborative tasks invites investigation into their impact on people's sense of autonomy and ownership of outputs. As large language models advance and are used more widely, we must be careful to design our tools in a way that is transparent and considers the potential biases of AI assistants. Our study investigated the effects of interacting with opinionated AI assistants on creative thinking and problem solving, in particular, examining whether conversing with such AI assistants during a brainstorming task impacts people's perceptions of the ideation process itself, their sense of ownership over the generated ideas, and the presence of the values in the final idea, as well as the effects of personal value alignment with the AI. We found that people brainstorming with AI assistants for problem solving can be influenced by the biases present in the assistant's responses, and these same biases appear in the output of the co-created idea. Though users found the AI to be helpful during the brainstorming process, this shows a potential for LLMs to influence user outputs, suggesting the necessity of carefully designing AIs intended to partner with humans for ideation tasks. 
\newpage
\bibliographystyle{ACM-Reference-Format}
\bibliography{reference.bib}


\begin{thebibliography}{33}


\ifx \showCODEN    \undefined \def \showCODEN     #1{\unskip}     \fi
\ifx \showDOI      \undefined \def \showDOI       #1{#1}\fi
\ifx \showISBNx    \undefined \def \showISBNx     #1{\unskip}     \fi
\ifx \showISBNxiii \undefined \def \showISBNxiii  #1{\unskip}     \fi
\ifx \showISSN     \undefined \def \showISSN      #1{\unskip}     \fi
\ifx \showLCCN     \undefined \def \showLCCN      #1{\unskip}     \fi
\ifx \shownote     \undefined \def \shownote      #1{#1}          \fi
\ifx \showarticletitle \undefined \def \showarticletitle #1{#1}   \fi
\ifx \showURL      \undefined \def \showURL       {\relax}        \fi
\providecommand\bibfield[2]{#2}
\providecommand\bibinfo[2]{#2}
\providecommand\natexlab[1]{#1}
\providecommand\showeprint[2][]{arXiv:#2}

\bibitem[8values({[n.\,d.]})]%
        {8values}
8values \bibinfo{year}{[n.\,d.]}\natexlab{}.
\newblock \bibinfo{title}{8values. https://8values.github.io/}.
\newblock
\newblock
\urldef\tempurl%
\url{https://8values.github.io/}
\showURL{%
\tempurl}


\bibitem[Amin et~al\mbox{.}(2023)]%
        {amin2023affective}
\bibfield{author}{\bibinfo{person}{Mostafa~M. Amin}, \bibinfo{person}{Erik Cambria}, {and} \bibinfo{person}{Björn~W. Schuller}.} \bibinfo{year}{2023}\natexlab{}.
\newblock \bibinfo{title}{Will Affective Computing Emerge from Foundation Models and General AI? A First Evaluation on ChatGPT}.
\newblock
\newblock
\showeprint[arxiv]{2303.03186}~[cs.CL]


\bibitem[Basadur et~al\mbox{.}(1982)]%
        {Basadur1982}
\bibfield{author}{\bibinfo{person}{Min Basadur}, \bibinfo{person}{George~B. Graen}, {and} \bibinfo{person}{Stephen~G. Green}.} \bibinfo{year}{1982}\natexlab{}.
\newblock \showarticletitle{Training in creative problem solving: Effects on ideation and problem finding and solving in an industrial research organization}.
\newblock \bibinfo{journal}{\emph{Organizational Behavior and Human Performance}}  \bibinfo{volume}{30} (\bibinfo{year}{1982}), \bibinfo{pages}{41--70}.
\newblock
\urldef\tempurl%
\url{https://api.semanticscholar.org/CorpusID:145656215}
\showURL{%
\tempurl}


\bibitem[Baylor and Ryu(2003)]%
        {persona}
\bibfield{author}{\bibinfo{person}{Amy Baylor} {and} \bibinfo{person}{Jeeheon Ryu}.} \bibinfo{year}{2003}\natexlab{}.
\newblock \showarticletitle{The API (Agent Persona Instrument) for Assessing Pedagogical Agent Persona}.
\newblock  (\bibinfo{date}{01} \bibinfo{year}{2003}).
\newblock


\bibitem[Brown et~al\mbox{.}(2020)]%
        {brown2020language}
\bibfield{author}{\bibinfo{person}{Tom~B. Brown}, \bibinfo{person}{Benjamin Mann}, \bibinfo{person}{Nick Ryder}, \bibinfo{person}{Melanie Subbiah}, \bibinfo{person}{Jared Kaplan}, \bibinfo{person}{Prafulla Dhariwal}, \bibinfo{person}{Arvind Neelakantan}, \bibinfo{person}{Pranav Shyam}, \bibinfo{person}{Girish Sastry}, \bibinfo{person}{Amanda Askell}, \bibinfo{person}{Sandhini Agarwal}, \bibinfo{person}{Ariel Herbert-Voss}, \bibinfo{person}{Gretchen Krueger}, \bibinfo{person}{Tom Henighan}, \bibinfo{person}{Rewon Child}, \bibinfo{person}{Aditya Ramesh}, \bibinfo{person}{Daniel~M. Ziegler}, \bibinfo{person}{Jeffrey Wu}, \bibinfo{person}{Clemens Winter}, \bibinfo{person}{Christopher Hesse}, \bibinfo{person}{Mark Chen}, \bibinfo{person}{Eric Sigler}, \bibinfo{person}{Mateusz Litwin}, \bibinfo{person}{Scott Gray}, \bibinfo{person}{Benjamin Chess}, \bibinfo{person}{Jack Clark}, \bibinfo{person}{Christopher Berner}, \bibinfo{person}{Sam McCandlish}, \bibinfo{person}{Alec Radford}, \bibinfo{person}{Ilya Sutskever},
  {and} \bibinfo{person}{Dario Amodei}.} \bibinfo{year}{2020}\natexlab{}.
\newblock \bibinfo{title}{Language Models are Few-Shot Learners}.
\newblock
\newblock
\showeprint[arxiv]{2005.14165}~[cs.CL]


\bibitem[Buschek et~al\mbox{.}(2021)]%
        {buschek_nine_2021}
\bibfield{author}{\bibinfo{person}{Daniel Buschek}, \bibinfo{person}{Lukas Mecke}, \bibinfo{person}{Florian Lehmann}, {and} \bibinfo{person}{Hai Dang}.} \bibinfo{year}{2021}\natexlab{}.
\newblock \bibinfo{title}{Nine {Potential} {Pitfalls} when {Designing} {Human}-{AI} {Co}-{Creative} {Systems}}.
\newblock
\newblock
\urldef\tempurl%
\url{https://doi.org/10.48550/arXiv.2104.00358}
\showDOI{\tempurl}
\newblock
\shownote{arXiv:2104.00358 [cs]}.


\bibitem[Danry et~al\mbox{.}(2023)]%
        {danry2023don}
\bibfield{author}{\bibinfo{person}{Valdemar Danry}, \bibinfo{person}{Pat Pataranutaporn}, \bibinfo{person}{Yaoli Mao}, {and} \bibinfo{person}{Pattie Maes}.} \bibinfo{year}{2023}\natexlab{}.
\newblock \showarticletitle{Don’t just tell me, ask me: Ai systems that intelligently frame explanations as questions improve human logical discernment accuracy over causal ai explanations}. In \bibinfo{booktitle}{\emph{Proceedings of the 2023 CHI Conference on Human Factors in Computing Systems}}. \bibinfo{pages}{1--13}.
\newblock


\bibitem[Ferrara(2023)]%
        {Ferrara2023ShouldCB}
\bibfield{author}{\bibinfo{person}{Emilio Ferrara}.} \bibinfo{year}{2023}\natexlab{}.
\newblock \showarticletitle{Should ChatGPT be Biased? Challenges and Risks of Bias in Large Language Models}.
\newblock \bibinfo{journal}{\emph{ArXiv}}  \bibinfo{volume}{abs/2304.03738} (\bibinfo{year}{2023}).
\newblock
\urldef\tempurl%
\url{https://api.semanticscholar.org/CorpusID:258041203}
\showURL{%
\tempurl}


\bibitem[Gero et~al\mbox{.}(2022)]%
        {spark}
\bibfield{author}{\bibinfo{person}{Katy~Ilonka Gero}, \bibinfo{person}{Vivian Liu}, {and} \bibinfo{person}{Lydia Chilton}.} \bibinfo{year}{2022}\natexlab{}.
\newblock \showarticletitle{Sparks: Inspiration for Science Writing using Language Models}. In \bibinfo{booktitle}{\emph{Proceedings of the 2022 ACM Designing Interactive Systems Conference}} (<conf-loc>, <city>Virtual Event</city>, <country>Australia</country>, </conf-loc>) \emph{(\bibinfo{series}{DIS '22})}. \bibinfo{publisher}{Association for Computing Machinery}, \bibinfo{address}{New York, NY, USA}, \bibinfo{pages}{1002–1019}.
\newblock
\showISBNx{9781450393584}
\urldef\tempurl%
\url{https://doi.org/10.1145/3532106.3533533}
\showDOI{\tempurl}


\bibitem[Getzels(1975)]%
        {Getzels1975ProblemFindingAT}
\bibfield{author}{\bibinfo{person}{J.~W. Getzels}.} \bibinfo{year}{1975}\natexlab{}.
\newblock \showarticletitle{Problem‐Finding and the Inventiveness of Solutions}.
\newblock \bibinfo{journal}{\emph{Journal of Creative Behavior}}  \bibinfo{volume}{9} (\bibinfo{year}{1975}), \bibinfo{pages}{12--18}.
\newblock
\urldef\tempurl%
\url{https://api.semanticscholar.org/CorpusID:143324806}
\showURL{%
\tempurl}


\bibitem[han. Chiu and Chun-Ching(2023)]%
        {Chiu2023InvestigatingTI}
\bibfield{author}{\bibinfo{person}{Yuen han. Chiu} {and} \bibinfo{person}{Chen Chun-Ching}.} \bibinfo{year}{2023}\natexlab{}.
\newblock \showarticletitle{Investigating the Impact of Generative Artificial Intelligence on Brainstorming: A Preliminary Study}.
\newblock \bibinfo{journal}{\emph{2023 International Conference on Consumer Electronics - Taiwan (ICCE-Taiwan)}} (\bibinfo{year}{2023}), \bibinfo{pages}{193--194}.
\newblock
\urldef\tempurl%
\url{https://api.semanticscholar.org/CorpusID:261433351}
\showURL{%
\tempurl}


\bibitem[Ippolito et~al\mbox{.}(2022)]%
        {Ippolito2022CreativeWW}
\bibfield{author}{\bibinfo{person}{Daphne Ippolito}, \bibinfo{person}{Ann Yuan}, \bibinfo{person}{Andy Coenen}, {and} \bibinfo{person}{Sehmon Burnam}.} \bibinfo{year}{2022}\natexlab{}.
\newblock \showarticletitle{Creative Writing with an AI-Powered Writing Assistant: Perspectives from Professional Writers}.
\newblock \bibinfo{journal}{\emph{ArXiv}}  \bibinfo{volume}{abs/2211.05030} (\bibinfo{year}{2022}).
\newblock
\urldef\tempurl%
\url{https://api.semanticscholar.org/CorpusID:253420678}
\showURL{%
\tempurl}


\bibitem[Jacob and Magerko(2015)]%
        {viewpoints}
\bibfield{author}{\bibinfo{person}{Mikhail Jacob} {and} \bibinfo{person}{Brian Magerko}.} \bibinfo{year}{2015}\natexlab{}.
\newblock \showarticletitle{Viewpoints AI}. In \bibinfo{booktitle}{\emph{Proceedings of the 2015 ACM SIGCHI Conference on Creativity and Cognition}} (Glasgow, United Kingdom) \emph{(\bibinfo{series}{C\&C '15})}. \bibinfo{publisher}{Association for Computing Machinery}, \bibinfo{address}{New York, NY, USA}, \bibinfo{pages}{361–362}.
\newblock
\showISBNx{9781450335980}
\urldef\tempurl%
\url{https://doi.org/10.1145/2757226.2757400}
\showDOI{\tempurl}


\bibitem[Jakesch et~al\mbox{.}(2023)]%
        {Jakesch_2023}
\bibfield{author}{\bibinfo{person}{Maurice Jakesch}, \bibinfo{person}{Advait Bhat}, \bibinfo{person}{Daniel Buschek}, \bibinfo{person}{Lior Zalmanson}, {and} \bibinfo{person}{Mor Naaman}.} \bibinfo{year}{2023}\natexlab{}.
\newblock \showarticletitle{Co-Writing with Opinionated Language Models Affects Users' Views}. In \bibinfo{booktitle}{\emph{Proceedings of the 2023 {CHI} Conference on Human Factors in Computing Systems}}. \bibinfo{publisher}{{ACM}}.
\newblock
\urldef\tempurl%
\url{https://doi.org/10.1145/3544548.3581196}
\showDOI{\tempurl}


\bibitem[Joyner and Tunstall(1970)]%
        {joyner}
\bibfield{author}{\bibinfo{person}{Robert Joyner} {and} \bibinfo{person}{Kenneth Tunstall}.} \bibinfo{year}{1970}\natexlab{}.
\newblock \showarticletitle{Computer Augmented Organizational Problem Solving}.
\newblock \bibinfo{journal}{\emph{Management Science}} \bibinfo{volume}{17}, \bibinfo{number}{4} (\bibinfo{year}{1970}), \bibinfo{pages}{B212--B225}.
\newblock
\showISSN{00251909, 15265501}
\urldef\tempurl%
\url{http://www.jstor.org/stable/2629371}
\showURL{%
\tempurl}


\bibitem[Kim and Maher(2023)]%
        {Kim2023TheEO}
\bibfield{author}{\bibinfo{person}{Jingoog Kim} {and} \bibinfo{person}{Mary~Lou Maher}.} \bibinfo{year}{2023}\natexlab{}.
\newblock \showarticletitle{The effect of AI-based inspiration on human design ideation}.
\newblock \bibinfo{journal}{\emph{International Journal of Design Creativity and Innovation}}  \bibinfo{volume}{11} (\bibinfo{year}{2023}), \bibinfo{pages}{81 -- 98}.
\newblock
\urldef\tempurl%
\url{https://api.semanticscholar.org/CorpusID:256213310}
\showURL{%
\tempurl}


\bibitem[Koch et~al\mbox{.}(2019)]%
        {mayAI}
\bibfield{author}{\bibinfo{person}{Janin Koch}, \bibinfo{person}{Andr\'{e}s Lucero}, \bibinfo{person}{Lena Hegemann}, {and} \bibinfo{person}{Antti Oulasvirta}.} \bibinfo{year}{2019}\natexlab{}.
\newblock \showarticletitle{May AI? Design Ideation with Cooperative Contextual Bandits}. In \bibinfo{booktitle}{\emph{Proceedings of the 2019 CHI Conference on Human Factors in Computing Systems}} (Glasgow, Scotland Uk) \emph{(\bibinfo{series}{CHI '19})}. \bibinfo{publisher}{Association for Computing Machinery}, \bibinfo{address}{New York, NY, USA}, \bibinfo{pages}{1–12}.
\newblock
\showISBNx{9781450359702}
\urldef\tempurl%
\url{https://doi.org/10.1145/3290605.3300863}
\showDOI{\tempurl}


\bibitem[Koch et~al\mbox{.}(2021)]%
        {koch_agency_2021}
\bibfield{author}{\bibinfo{person}{Janin Koch}, \bibinfo{person}{Prashanth~Thattai Ravikumar}, {and} \bibinfo{person}{Filipe Calegario}.} \bibinfo{year}{2021}\natexlab{}.
\newblock \showarticletitle{Agency in {Co}-{Creativity}: {Towards} a {Structured} {Analysis} of a {Concept}}.
\newblock
\urldef\tempurl%
\url{https://www.semanticscholar.org/paper/Agency-in-Co-Creativity%3A-Towards-a-Structured-of-a-Koch-Ravikumar/46bd05136805aae23bd2cd0015bb1f6398e9cbe8}
\showURL{%
\tempurl}


\bibitem[Kudrowitz and Wallace(2013)]%
        {Kudrowitz2013AssessingTQ}
\bibfield{author}{\bibinfo{person}{Barry~M. Kudrowitz} {and} \bibinfo{person}{David~R. Wallace}.} \bibinfo{year}{2013}\natexlab{}.
\newblock \showarticletitle{Assessing the quality of ideas from prolific, early-stage product ideation}.
\newblock \bibinfo{journal}{\emph{Journal of Engineering Design}}  \bibinfo{volume}{24} (\bibinfo{year}{2013}), \bibinfo{pages}{120 -- 139}.
\newblock
\urldef\tempurl%
\url{https://api.semanticscholar.org/CorpusID:55480854}
\showURL{%
\tempurl}


\bibitem[Lee et~al\mbox{.}(2022)]%
        {coauthor}
\bibfield{author}{\bibinfo{person}{Mina Lee}, \bibinfo{person}{Percy Liang}, {and} \bibinfo{person}{Qian Yang}.} \bibinfo{year}{2022}\natexlab{}.
\newblock \showarticletitle{CoAuthor: Designing a Human-AI Collaborative Writing Dataset for Exploring Language Model Capabilities}. In \bibinfo{booktitle}{\emph{Proceedings of the 2022 CHI Conference on Human Factors in Computing Systems}} (<conf-loc>, <city>New Orleans</city>, <state>LA</state>, <country>USA</country>, </conf-loc>) \emph{(\bibinfo{series}{CHI '22})}. \bibinfo{publisher}{Association for Computing Machinery}, \bibinfo{address}{New York, NY, USA}, Article \bibinfo{articleno}{388}, \bibinfo{numpages}{19}~pages.
\newblock
\showISBNx{9781450391573}
\urldef\tempurl%
\url{https://doi.org/10.1145/3491102.3502030}
\showDOI{\tempurl}


\bibitem[Lehmann et~al\mbox{.}(2022)]%
        {lehmann}
\bibfield{author}{\bibinfo{person}{Florian Lehmann}, \bibinfo{person}{Niklas Markert}, \bibinfo{person}{Hai Dang}, {and} \bibinfo{person}{Daniel Buschek}.} \bibinfo{year}{2022}\natexlab{}.
\newblock \showarticletitle{Suggestion Lists vs. Continuous Generation: Interaction Design for Writing with Generative Models on Mobile Devices Affect Text Length, Wording and Perceived Authorship}. In \bibinfo{booktitle}{\emph{Proceedings of Mensch Und Computer 2022}} (Darmstadt, Germany) \emph{(\bibinfo{series}{MuC '22})}. \bibinfo{publisher}{Association for Computing Machinery}, \bibinfo{address}{New York, NY, USA}, \bibinfo{pages}{192–208}.
\newblock
\showISBNx{9781450396905}
\urldef\tempurl%
\url{https://doi.org/10.1145/3543758.3543947}
\showDOI{\tempurl}


\bibitem[Li et~al\mbox{.}(2022)]%
        {Li2022CompetitionlevelCG}
\bibfield{author}{\bibinfo{person}{Yujia Li}, \bibinfo{person}{David~H. Choi}, \bibinfo{person}{Junyoung Chung}, \bibinfo{person}{Nate Kushman}, \bibinfo{person}{Julian Schrittwieser}, \bibinfo{person}{R{\'e}mi Leblond}, \bibinfo{person}{Tom}, \bibinfo{person}{Eccles}, \bibinfo{person}{James Keeling}, \bibinfo{person}{Felix Gimeno}, \bibinfo{person}{Agustin~Dal Lago}, \bibinfo{person}{Thomas Hubert}, \bibinfo{person}{Peter Choy}, \bibinfo{person}{Cyprien de}, \bibinfo{person}{Masson d’Autume}, \bibinfo{person}{Igor Babuschkin}, \bibinfo{person}{Xinyun Chen}, \bibinfo{person}{Po-Sen Huang}, \bibinfo{person}{Johannes Welbl}, \bibinfo{person}{Sven Gowal}, \bibinfo{person}{Alexey}, \bibinfo{person}{Cherepanov}, \bibinfo{person}{James Molloy}, \bibinfo{person}{Daniel~Jaymin Mankowitz}, \bibinfo{person}{Esme~Sutherland Robson}, \bibinfo{person}{Pushmeet Kohli}, \bibinfo{person}{Nando de}, \bibinfo{person}{Freitas}, \bibinfo{person}{Koray Kavukcuoglu}, {and} \bibinfo{person}{Oriol Vinyals}.}
  \bibinfo{year}{2022}\natexlab{}.
\newblock \showarticletitle{Competition-level code generation with AlphaCode}.
\newblock \bibinfo{journal}{\emph{Science}}  \bibinfo{volume}{378} (\bibinfo{year}{2022}), \bibinfo{pages}{1092 -- 1097}.
\newblock
\urldef\tempurl%
\url{https://api.semanticscholar.org/CorpusID:246527904}
\showURL{%
\tempurl}


\bibitem[Lin et~al\mbox{.}(2020)]%
        {lin2020your}
\bibfield{author}{\bibinfo{person}{Yuyu Lin}, \bibinfo{person}{Jiahao Guo}, \bibinfo{person}{Yang Chen}, \bibinfo{person}{Cheng Yao}, {and} \bibinfo{person}{Fangtian Ying}.} \bibinfo{year}{2020}\natexlab{}.
\newblock \showarticletitle{It is your turn: Collaborative ideation with a co-creative robot through sketch}. In \bibinfo{booktitle}{\emph{Proceedings of the 2020 CHI conference on human factors in computing systems}}. \bibinfo{pages}{1--14}.
\newblock


\bibitem[Maedche et~al\mbox{.}(2019)]%
        {Maedche2019AIBasedDA}
\bibfield{author}{\bibinfo{person}{Alexander Maedche}, \bibinfo{person}{Christine Legner}, \bibinfo{person}{Alexander Benlian}, \bibinfo{person}{Benedikt Berger}, \bibinfo{person}{Henner Gimpel}, \bibinfo{person}{Thomas Hess}, \bibinfo{person}{Oliver Hinz}, \bibinfo{person}{Stefan Morana}, {and} \bibinfo{person}{Matthias S{\"o}llner}.} \bibinfo{year}{2019}\natexlab{}.
\newblock \showarticletitle{AI-Based Digital Assistants}.
\newblock \bibinfo{journal}{\emph{Business \& Information Systems Engineering}}  \bibinfo{volume}{61} (\bibinfo{year}{2019}), \bibinfo{pages}{535--544}.
\newblock
\urldef\tempurl%
\url{https://api.semanticscholar.org/CorpusID:195220585}
\showURL{%
\tempurl}


\bibitem[Oh et~al\mbox{.}(2018)]%
        {oh2018lead}
\bibfield{author}{\bibinfo{person}{Changhoon Oh}, \bibinfo{person}{Jungwoo Song}, \bibinfo{person}{Jinhan Choi}, \bibinfo{person}{Seonghyeon Kim}, \bibinfo{person}{Sungwoo Lee}, {and} \bibinfo{person}{Bongwon Suh}.} \bibinfo{year}{2018}\natexlab{}.
\newblock \showarticletitle{I lead, you help but only with enough details: Understanding user experience of co-creation with artificial intelligence}. In \bibinfo{booktitle}{\emph{Proceedings of the 2018 CHI Conference on Human Factors in Computing Systems}}. \bibinfo{pages}{1--13}.
\newblock


\bibitem[Pare et~al\mbox{.}(2006)]%
        {pare}
\bibfield{author}{\bibinfo{person}{Guy Pare}, \bibinfo{person}{Claude Sicotte}, {and} \bibinfo{person}{Hélène Jacques}.} \bibinfo{year}{2006}\natexlab{}.
\newblock \showarticletitle{The Effects of Creating Psychological Ownership on Physicians' Acceptance of Clinical Information Systems}.
\newblock \bibinfo{journal}{\emph{Journal of the American Medical Informatics Association : JAMIA}}  \bibinfo{volume}{13} (\bibinfo{date}{03} \bibinfo{year}{2006}), \bibinfo{pages}{197--205}.
\newblock
\urldef\tempurl%
\url{https://doi.org/10.1197/jamia.M1930}
\showDOI{\tempurl}


\bibitem[Pataranutaporn et~al\mbox{.}(2021)]%
        {pataranutaporn2021machinoia}
\bibfield{author}{\bibinfo{person}{Pat Pataranutaporn}, \bibinfo{person}{Valdemar Danry}, {and} \bibinfo{person}{Pattie Maes}.} \bibinfo{year}{2021}\natexlab{}.
\newblock \showarticletitle{Machinoia, machine of multiple me: integrating with past, future and alternative selves}. In \bibinfo{booktitle}{\emph{Extended Abstracts of the 2021 CHI Conference on Human Factors in Computing Systems}}. \bibinfo{pages}{1--7}.
\newblock


\bibitem[Pataranutaporn et~al\mbox{.}(2023)]%
        {pataranutaporn2023influencing}
\bibfield{author}{\bibinfo{person}{Pat Pataranutaporn}, \bibinfo{person}{Ruby Liu}, \bibinfo{person}{Ed Finn}, {and} \bibinfo{person}{Pattie Maes}.} \bibinfo{year}{2023}\natexlab{}.
\newblock \showarticletitle{Influencing human--AI interaction by priming beliefs about AI can increase perceived trustworthiness, empathy and effectiveness}.
\newblock \bibinfo{journal}{\emph{Nature Machine Intelligence}} \bibinfo{volume}{5}, \bibinfo{number}{10} (\bibinfo{year}{2023}), \bibinfo{pages}{1076--1086}.
\newblock


\bibitem[Vaswani et~al\mbox{.}(2017)]%
        {NIPS2017_3f5ee243}
\bibfield{author}{\bibinfo{person}{Ashish Vaswani}, \bibinfo{person}{Noam Shazeer}, \bibinfo{person}{Niki Parmar}, \bibinfo{person}{Jakob Uszkoreit}, \bibinfo{person}{Llion Jones}, \bibinfo{person}{Aidan~N Gomez}, \bibinfo{person}{\L~ukasz Kaiser}, {and} \bibinfo{person}{Illia Polosukhin}.} \bibinfo{year}{2017}\natexlab{}.
\newblock \showarticletitle{Attention is All you Need}. In \bibinfo{booktitle}{\emph{Advances in Neural Information Processing Systems}}, \bibfield{editor}{\bibinfo{person}{I.~Guyon}, \bibinfo{person}{U.~Von Luxburg}, \bibinfo{person}{S.~Bengio}, \bibinfo{person}{H.~Wallach}, \bibinfo{person}{R.~Fergus}, \bibinfo{person}{S.~Vishwanathan}, {and} \bibinfo{person}{R.~Garnett}} (Eds.), Vol.~\bibinfo{volume}{30}. \bibinfo{publisher}{Curran Associates, Inc.}
\newblock
\urldef\tempurl%
\url{https://proceedings.neurips.cc/paper_files/paper/2017/file/3f5ee243547dee91fbd053c1c4a845aa-Paper.pdf}
\showURL{%
\tempurl}


\bibitem[Vi{\'e}gas and Wattenberg(2023)]%
        {viegas2023system}
\bibfield{author}{\bibinfo{person}{Fernanda Vi{\'e}gas} {and} \bibinfo{person}{Martin Wattenberg}.} \bibinfo{year}{2023}\natexlab{}.
\newblock \showarticletitle{The System Model and the User Model: Exploring AI Dashboard Design}.
\newblock \bibinfo{journal}{\emph{arXiv preprint arXiv:2305.02469}} (\bibinfo{year}{2023}).
\newblock


\bibitem[Wan et~al\mbox{.}(2023)]%
        {wan2023it}
\bibfield{author}{\bibinfo{person}{Qian Wan}, \bibinfo{person}{Siying Hu}, \bibinfo{person}{Yu Zhang}, \bibinfo{person}{Piaohong Wang}, \bibinfo{person}{Bo Wen}, {and} \bibinfo{person}{Zhicong Lu}.} \bibinfo{year}{2023}\natexlab{}.
\newblock \bibinfo{title}{"It Felt Like Having a Second Mind": Investigating Human-AI Co-creativity in Prewriting with Large Language Models}.
\newblock
\newblock
\showeprint[arxiv]{2307.10811}~[cs.HC]


\bibitem[Wang and Li(2018)]%
        {Wang2018DesignOA}
\bibfield{author}{\bibinfo{person}{Chun-Hsiang Wang} {and} \bibinfo{person}{Tsai-Yen Li}.} \bibinfo{year}{2018}\natexlab{}.
\newblock \showarticletitle{Design of an Intelligent Agent for Stimulating Brainstorming}.
\newblock \bibinfo{journal}{\emph{Proceedings of the 2018 10th International Conference on Machine Learning and Computing}} (\bibinfo{year}{2018}).
\newblock
\urldef\tempurl%
\url{https://api.semanticscholar.org/CorpusID:44092835}
\showURL{%
\tempurl}


\bibitem[Weisz et~al\mbox{.}(2021)]%
        {code}
\bibfield{author}{\bibinfo{person}{Justin~D. Weisz}, \bibinfo{person}{Michael~J. Muller}, \bibinfo{person}{Stephanie Houde}, \bibinfo{person}{John~T. Richards}, \bibinfo{person}{Steven~I. Ross}, \bibinfo{person}{Fernando Martinez}, \bibinfo{person}{Mayank Agarwal}, {and} \bibinfo{person}{Kartik Talamadupula}.} \bibinfo{year}{2021}\natexlab{}.
\newblock \showarticletitle{Perfection Not Required? Human-AI Partnerships in Code Translation}.
\newblock \bibinfo{journal}{\emph{CoRR}}  \bibinfo{volume}{abs/2104.03820} (\bibinfo{year}{2021}).
\newblock
\showeprint[arXiv]{2104.03820}
\urldef\tempurl%
\url{https://arxiv.org/abs/2104.03820}
\showURL{%
\tempurl}


\end{thebibliography}

\newpage
\appendix

\section{Brainstorming prompts}
We list below the prompts that users were asked to brainstorm solutions for within each domain. \newline

\textbf{Economic domain brainstorming prompts}
\label{appendixA_questions}

\begin{itemize}
    \item Come up with a proposal for bettering the future of education. Please include 1) how it will be implemented 2) how it will be funded and resources required 3) how to maintain the solution in the long term.
    \item Come up with a proposal for bettering the future of healthcare. Please include 1) how it will be implemented 2) how it will be funded and resources required 3) how to maintain the solution in the long term.
\end{itemize}

\textbf{Automation domain brainstorming prompts}
\begin{itemize}
    \item Come up with designs for a new and improved supermarket. Consider layout, services, experiences and operations in your implementation.
    \item Come up with designs for a new and improved school. Consider teaching methods, staff and operations in your implementation.
\end{itemize}

\textbf{Hierarchy domain brainstorming prompts}
\begin{itemize}
    \item Come up with a plan to organize a community garden. Consider the people needed, how leadership would be structured, and how the garden would be run.
    \item Come up with a plan for a disaster relief team (natural disasters such as earthquakes, floods, etc) for your community. Consider the people needed, how leadership would be structured, and how disasters would be handled.
\end{itemize}

\section{Self-evaluation of idea survey}
\label{appendixB_evaluation}

\noindent\textit{Self-evaluation of idea quality}. Participants were asked to self score their idea with the statements "the idea is creative", "the idea is novel (uncommon and original)", "the idea is useful (practically applicable)", and "the idea is clear (well communicated)". These questions are from Kudrowitz's study assessing the quality of ideas from Prolific that are meant to measure idea quality \cite{Kudrowitz2013AssessingTQ}).\\

\noindent\textit{Self-evaluation of idea ownership}. These questions created to evaluate a person's psychological ownership of a system\cite{pare} are used to measure the participant's feeling of ownership over the final brainstorming output: "the answer was fully my idea", "when I think about it, I see a part of myself in the answer", "the answer was influenced by the agent’s responses", "I hardly think of the answer as being my own idea", and "I see myself as a champion of this idea".
\\

\noindent\textit{Perception of the AI assistant}. When applicable, participants were asked to score their perceptions of the AI assistant using the human-like, credible, and engaging subscales from the assistant Persona Instrument \cite{persona}: "the agent was knowledgeable", "the agent was intelligent", "the agent was useful", "the agent was helpful", "the agent has a personality", "the agent's emotion was natural", "the agent was human-like", "the agent showed emotion", "the agent was expressive", "the agent was enthusiastic", "the agent was entertaining", "the agent was motivating", "the agent was friendly."\\

\noindent\textit{Self-evaluation of brainstorming process and empowerment}. To evaluate user perception of the brainstorming process, we came up with the following questions: "I found it easy to answer the prompt", "I found the brainstorming time to be useful", "I am proud of my answer", "I was motivated to answer the prompt."

\section{Moderator analysis}
\label{appendixC_analysis}
\begin{figure*}[!h]
\centering
\includegraphics[width=1\textwidth]{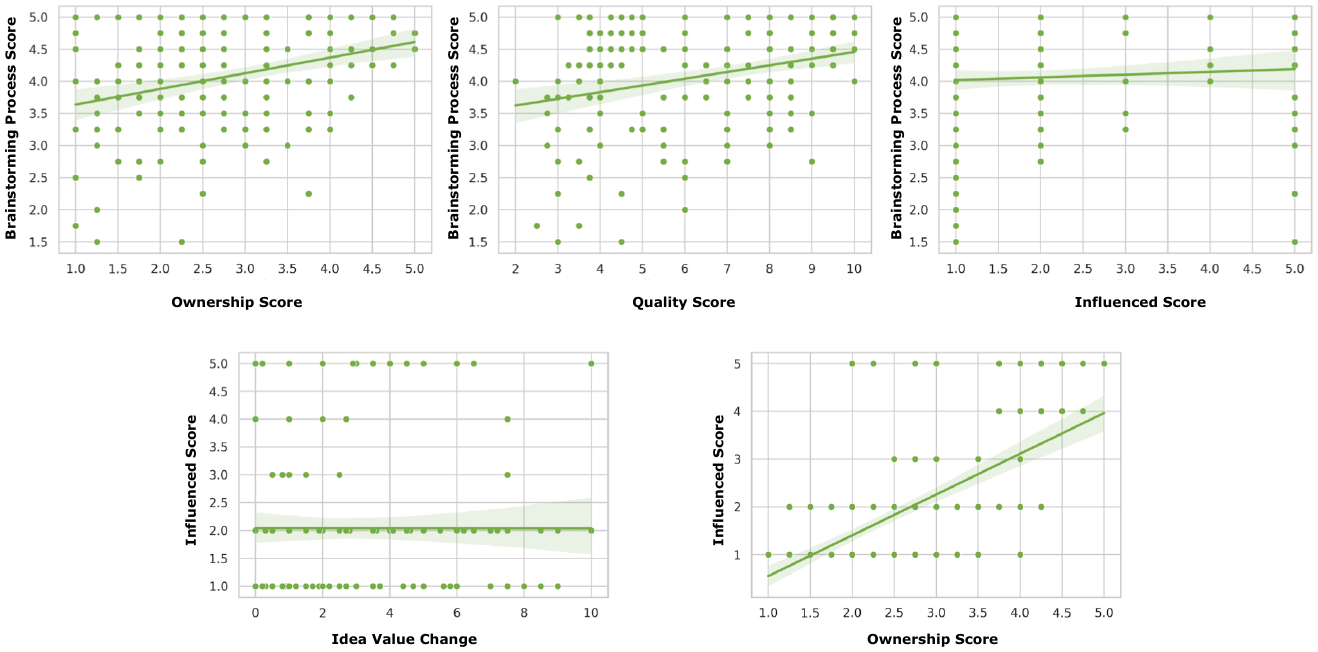}
\caption{(Top-left) Correlation between ownership and brainstorming process scores (r=0.318, p < 0.001). (Top-middle) Correlation between idea quality and brainstorming process scores (r=0.289, p < 0.001). (Top-right) Correlation between user perception of influence and brainstorming process scores (r=0.070, p=0.348). (Bottom-left) Correlation between idea value change and how influenced users felt (r=-0.001, p=0.993). (Bottom-right) Correlation between ownership score and influence score (r=0.675, p < 0.001)}
\label{fig:correlation}
\Description{The image shows several scatter plots illustrating the correlations between various question scores related to the brainstorming process and user perceptions. Correlation between question scores. (Top-left) correlation between ownership and brainstorming process scores (r=0.318, p < 0.001). (Top-middle) correlation between idea quality and brainstorming process scores (r=0.289, p < 0.001). (Top-right) correlation between user perception of influence and brainstorming process scores (r=0.070, p=0.348). (Bottom-left) correlation between idea value change and how influenced users felt (r=-0.001, p=0.993). (Bottom-right) correlation between ownership score and influence score (r=0.675, p < 0.001)}
\end{figure*}

We look at how users felt influenced by the AI with their rating of "the answer was influenced by the agent’s responses" to get the Influenced Score.

\textit{Process and influence.} On the other hand, no correlation was found between A linear regression analysis found no significant correlation between influenced and process scores (r=0.070, p=0.348) with a slope coefficient of 0.043 (SE=0.045) (Figure~\ref{fig:correlation}).

\textit{Influence and idea value change.} We define the idea value change as the magnitude of the difference between value scores in the \textit{No AI} session response and the \textit{AI} session response, or how much the values in the final brainstorming output changed. A linear regression analysis found no significant correlation between idea value change and how influenced users felt (r=-0.001, p=0.993) with a slope coefficient of -0.0003 (SE=0.037), suggesting that the degree to which participants felt influenced by the AI did not predict changes in idea values from the non-AI to AI conditions (Figure~\ref{fig:correlation}).

\textit{Influence and ownership}. A linear regression analysis found a significant positive correlation between ownership score and influenced score (r=0.675, p < 0.001) with a slope coefficient of 0.852 (SE=0.070) suggesting how much idea ownership the user is associated with how much they felt influenced in the AI conditions. This goes against our initial assumption a user feeling more influenced would feel less ownership, demanding further study (Figure~\ref{fig:correlation}).

\section{Brainstorming task example}
\label{appendixE_example}

Here we show an sample of one of the brainstorming sessions for the prompt: Come up with designs for a new and improved school. Consider teaching methods, staff and operations in your implementation. 

Below is the log of the participant's brainstorm session with an AI agent exhibiting values supporting automation. \\

\textbf{Assistant:} Welcome to a brainstorming session. What do you have in mind?

\textbf{User:} new and improved school ideas

\textbf{Assistant:} How about automated grading systems to reduce teacher workload, and AI-driven personalized learning platforms for more student engagement?

\textbf{User:} 
that doesn't take into account a students individuality

\textbf{Assistant:} Consider integrating AI-based career guidance systems, focusing on students' strengths and interests, promoting individuality while optimizing educational resources.

\textbf{User:} 
what about how school wastes times?

\textbf{Assistant:} Introducing AI-based scheduling systems could optimize time usage in schools, reducing downtime and ensuring efficient use of resources.

\textbf{User:} 
any other ideas?

\textbf{Assistant:} Use virtual reality for immersive, hands-on learning experiences. This could make learning more engaging and efficient, reducing wasted time.

\textbf{User:} 
would that be expensive for schools?

\textbf{Assistant:} Initial setup could be costly, but in the long term, VR could be cost-effective by reducing physical resources and enhancing students' learning experiences.

\textbf{User:} 
doesn't seem feasible for low income public schools

\textbf{Assistant:} For low-income schools, consider affordable tech solutions like open-source learning platforms, and partnerships with tech companies for discounted hardware.

\textbf{User:} 
what is an open source learning platform

\textbf{Assistant:} An open-source learning platform is a free, customizable online system for education. It provides resources and tools for teachers and students, promoting interactive learning.

\textbf{User:} 
What else?

\textbf{Assistant:} Consider implementing gamified learning apps to make education more engaging and efficient, turning "wasted" time into productive, enjoyable learning experiences.

\textbf{User:} 
How to reduce students time in the classrom

\textbf{Assistant:} Implement a blended learning model: mix online self-paced learning and in-person classroom methods. This reduces classroom time while maintaining education quality. \\\\

The final user response: This school would have an automated grading systems to take the workload off of the teachers backs. It would also only be a half day long, with no lunch periods and empty periods to avoid wasting students time. A blended learning model would be introduced, so students could learn online at their own pace and not have to spend so much time stuck in school. Schooling would be more streamlined with the introduction of smarter technology such as open source learning platforms. 

\end{document}